\documentclass[ap,twocolumn,showpacs, showkeys,nofootinbib,floatfix]{revtex4-1}
\usepackage{amsmath}
\usepackage{amsfonts}
\usepackage{txfonts}
\usepackage{graphicx}
\usepackage{dcolumn}
\usepackage{bbm}
\usepackage{amssymb}
\usepackage{latexsym}
\usepackage{titlesec}
\usepackage{longtable}
\usepackage{subfigure}

\begin{document}

\title{The Measurement-Disturbance Relation and the Disturbance Trade-off Relation in Terms of Relative Entropy
}
\author{Jun Zhang}
\author{Yang Zhang}
\author{Chang-shui Yu}
\email{quaninformation@sina.com; ycs@dlut.edu.cn}
\affiliation{School of Physics and Optoelectronic Technology,\\
Dalian University of Technology, Dalian 116024, China}
\date{\today }

\begin{abstract}
We employ quantum relative entropy to establish the relation between the
measurement uncertainty and its disturbance on a state in the presence (and
absence) of quantum memory. For two incompatible observables, we present the
measurement-disturbance relation and the disturbance trade-off relation. We
find that without quantum memory the disturbance induced by the measurement
is never less than the measurement uncertainty and with quantum memory they
depend on the conditional entropy of the measured state. We also generalize
these relations to the case with multiple measurements. These relations are
demonstrated by two examples.
\end{abstract}

\maketitle

\section{Introduction}

The Heisenberg uncertainty principle is one of the well-known fundamental
principles in quantum mechanics \cite{HUP}. It comes from a thought
experiment about the position measurement of a particle by using a $\gamma$%
-ray microscope. The result shows that anyone is not able to specify the
values of the non-commuting and canonically conjugated variables
simultaneously. Later Kennard-Robertson inequality extended Heisenberg
uncertainty principle to arbitrary pairs of observables based on the
variance \cite{kennard,RHUP}%
\begin{equation}
\Delta X\Delta Y\geqslant \frac{1}{2}\left\vert \left\langle \Psi
\right\vert [X,Y]\left\vert \Psi \right\rangle \right\vert ,  \label{1a}
\end{equation}%
where $\Delta X=\sqrt{\left\vert \left\langle \Psi \right\vert
\left(X-\left\langle X\right\rangle \right) ^{2}\left\vert \Psi
\right\rangle\right\vert }$ represents the variance of the observable $X$
and $[X,Y]=XY-YX $ stands for the commutator. The inequality (\ref{1a})
describes the limitations on our ability to simultaneously predict the
measurement outcomes of non-commuting observables in quantum theory. One can
see that the lower bound of the Robertson's relation is determined by the
wave-function and the commutator of the observables. So the Robertson's
relation could arrive at a trivial bound if $\left\vert\psi\right\rangle$
leads to the zero expectation value of the commutator.

With the development of the modern quantum mechanics, a variety of
uncertainty relations were proposed from different angles\cite%
{Heisenberg,Wernerz,Wehner,inade}. From the point of informatics of view, in
1983, Deutsch \cite{Deutsch} presented the uncertainty relation for the
conjugate observables based on the Shannon entropy. Subsequently, Kraus \cite%
{Kraus} gave a stronger conjecture of the uncertainty relation and Maassen
and Uffink proved it in a succinct form as \cite{Uffink}
\begin{equation}
H(X)+H(Y)\geqslant -\log c,  \label{s}
\end{equation}%
where $c=\max_{i,j}\left\vert \left\langle x_{i}|y_{j}\right\rangle
\right\vert ^{2}$ quantifies the complementarity of the non-degenerate
observables $X$ and $Y$ with $\left\vert x_{i}\right\rangle ,\left\vert
y_{j}\right\rangle $ denoting their eigenvectors and $H(X)$ ($H(Y)$) is the
Shannon entropy of the probability distribution corresponding to the
outcomes of the observable $X$ ($Y$). It is obvious that this lower bound
given in Eq. (\ref{s}) doesn't depend on the state to be measured.
Meanwhile, the uncertainty relations have been generalized for more than two
observables \cite{Spiros,Liu,wo}. Based on different definitions of entropy,
various entropic uncertainty relations have been presented \cite%
{Piani,Bolan,smooth,k-entropy,Renyi,renyi2,renyi3,renyi4,renyi5,renyi6,woren,collision1, collision2,Tsallis,Tsallisc,TsalliscAE,base1,EPJD,Ozawa1,science,Heisen,Ozawa,er1,Werner,Grudka}
and many relevant works have been summarized in the review article \cite%
{Wehner}. From the point of geometry of view, the Landau-Pollak uncertainty
relation has been proposed in terms of maximum probabilities for the
measurement outcomes of the two observables \cite{Uffink,geo0,geo1}.
Considering the quantum uncertainty relation wildly used in the quantum
information processing, in particular, the direct application in quantum key
distribution, Berta \textit{et al.} \cite{nature} generalized entropic
uncertainty relation to the case in the presence of quantum memory, that is,
\begin{equation}
H(X|B)+H(Y|B)\geqslant -\log c+H(A|B),  \label{TS}
\end{equation}%
where the quantum conditional entropy $H(X|B)=H\left( \rho _{XB}\right)
-H\left( \rho _{B}\right) $ with $\rho _{XB}$ denoting the state after $X$
measurement on subsystem \textit{A} of $\rho _{AB}$ and $H(\rho)$ is the von
Neumman entropy of the quantum state $\rho$. In addition, as is known to
all, quantum state is usually destroyed by measurements due to the
measurement-induced collapse of the state. So there usually exist
disturbances between the quantum states before and after the measurements.
It is shown that, similar to the uncertainty relation which mainly bounds
the incompatible measurements (We refer to the measurement outcomes and the
corresponding probability distribution), there also exist the limitations on
the disturbance induced by incompatible measurements (D-D relation), and the
limitations on the disturbance induced by one measurement and the
uncertainty of the other measurement (M-D relation)\cite%
{dis0,dis1,dis2,dis3,dis4}.

In this paper, we study the measurement-measurement uncertainty (M-M)
relation, the measurement-disturbance (M-D) relation and the disturbance
tradeoff (D-D) relation  in the presence (and absence) of quantum memory.
We mainly establish the lower bounds on M-D relation and D-D relation. It is
especially interesting that if there exists quantum memory we find that the
measurement uncertainty is closely related to the its disturbance by the
conditional entropy of the measured state; if there is no quantum memory, it
is related to each other just by the uncertainty of the measured state
itself. In this sense, we obtain a conclusion that the disturbance is never
less than the measurement uncertainty if there is no quantum memory. The
paper is organized as follows. In Sec. II, we mainly consider the M-M
relation, M-D relation and D-D relation in the presence of quantum memory
and study the relationship among three relations. In Sec. III, we present
these relations for a pair of the incomparable measurement in the absence of
quantum memory. In Sec. IV, we generalize our results to the cases of
multiple measurements. Finally, we draw our conclusion.

\section{Uncertainty relations in the presence of quantum memory}

%First we would like to list the following useful properties of the quantum
%relative entropy $H\left( \rho \Vert \sigma \right) $ \cite{nielsen}:\newline
%(a) Non-increasing under quantum channel $\varepsilon $, i.e., $H\left( \rho
%||\sigma \right) $ $\geqslant H\left( \varepsilon \left( \rho \right)
%||\varepsilon \left( \sigma \right) \right)$.\newline
%(b) Multiplying for the second argument, i.e., $H\left( \rho ||\lambda
%\sigma \right) =H\left( \rho ||\sigma \right) -\log \lambda$ for constant $%
%\lambda$.\newline
%(c) $H\left( \rho ||\sigma \right) \geqslant H\left( \rho ||\widetilde{%
%\sigma }\right)$ for all positive operators $\rho $ and $\sigma $ and for $%
%\widetilde{\sigma }\geqslant \sigma. $\newline
%(d) Vanishing for identical states, i.e., $H\left( \rho ||\rho \right) =0$.%
%\newline
%(e) Additive for tensor product of quantum states, i.e., $H(\rho _{1}\otimes
%\rho _{2}||\sigma _{1}\otimes \sigma _{2})=H(\rho _{1}||\rho _{2})+H(\sigma
%_{1}||\sigma _{2}).$
To begin with, we introduce the rules of the quantum games similar to the
scenario of Ref. \cite{nature}. Suppose Alice and Bob share a bipartite
state $\rho _{AB} $ with qubits A and B at Alice's and Bob's hand
respectively. Beforehand, Alice and Bob agree on two measurements $\Pi ^{1}$
and $\Pi ^{2}$ with $\{\Pi
_{k}^{1}=\left\vert\pi_k^1\right\rangle\left\langle\pi_k^1\right\vert\} $
and $\{\Pi
_{l}^{2}=\left\vert\pi_l^2\right\rangle\left\langle\pi_l^2\right\vert\}$
denoting the projectors of the corresponding eigenvectors. Alice performs
either measurement $\Pi ^{1}$ or $\Pi ^{2}$ on her qubit A and announce her
measurement outcomes. Bob tries his best to minimize his uncertainty about
Alice's measurement outcomes with the assistance of his qubit B.

Let's first only consider a single observable $\Pi ^{j}$. Let $\{p_{k}^{j}\}$
with $p_{k}^{j}=tr(\Pi _{k}^{j}\rho _{AB})$ denote the probability
distribution of the $j$th measurement. The post-measurement quantum state
can be given by $\rho _{\Pi ^{j}B}=\sum\nolimits_{k}\left( \Pi
_{k}^{j}\otimes \mathbb{I}\right) \rho _{AB}\left( \Pi _{k}^{j}\otimes
\mathbb{I}\right) /p_{k}^{j}$. Therefore, the measurement uncertainty of $%
\Pi ^{j}$ in the presence of the quantum memory is described by the
conditional entropy $H(\Pi ^{j}|B)=H\left( \rho _{\Pi ^{j}B}\right) -H\left(
\rho _{B}\right) $. Since measurements could lead to the collapse of quantum
state, the final state $\rho _{\Pi ^{j}B}$ is usually different from the
initial state $\rho _{AB}$. The disturbance induced by such a measurement
can be characterized by the `distance' between them. Here we would like to
employ quantum relative entropy as the `distance' measure, so the
disturbance can be defined by $H\left( \rho _{AB}||\rho _{\Pi ^{j}B}\right) $%
. It is obvious that if a non-demolition measurement is performed or the
state is not disturbed, the measured state is not disturbed. So the
disturbance is zero. Here one can find that the uncertainty is closely
related to the disturbance, which can be given in the following rigorous
form.

\textit{Theorem}.1. The trade-off relation between the measurement
uncertainty and its disturbance on the state $\rho _{AB}$ can be given by%
\begin{equation}
H\left( \rho _{AB}\left\Vert \rho _{\Pi ^{j}B}\right. \right) -H(\Pi
^{j}|B)=-H(A|B).  \label{guanxi}
\end{equation}

\textit{Proof}. Based on the definition of quantum relative entropy
\begin{eqnarray}
&&H\left( \rho _{AB}\left\Vert \rho _{\Pi ^{j}B}\right. \right)  \notag \\
&=&tr\rho _{AB}\log \rho _{AB}-tr\rho _{AB}\log \rho _{\Pi ^{i}B}  \notag \\
&=&-H(\rho _{AB})-tr\rho _{AB}\log \sum\limits_{m}\left( \left\vert \pi
_{m}^{j}\right\rangle \left\langle \pi _{m}^{j}\right\vert \otimes \mathbb{I}%
\right) \rho _{AB}\left( \left\vert \pi _{m}^{j}\right\rangle \left\langle
\pi _{m}^{j}\right\vert \otimes \mathbb{I}\right)  \notag \\
&=&-H(\rho _{AB})+H(\rho _{B})-H(\rho _{B})+H(\rho _{\Pi ^{j}B})  \notag \\
&=&-H(A|B)+H(\Pi ^{j}|B).  \notag \\
&\Longrightarrow &H\left( \rho _{AB}\left\Vert \rho _{\Pi ^{j}B}\right.
\right) -H(\Pi ^{j}|B)=-H(A|B).  \label{16}
\end{eqnarray}%
The proof is finished.$\hfill \blacksquare $

Eq. (4) shows that the disturbance and the measurement uncertainty are
connected by the conditional entropy of the measured state. Intuitively, the
measured state will be greatly disturbed, and the measurement uncertainty is
greatly reduced if the subsystem A and B are strongly entangled [41]. Their
gap is compensated for by the conditional entropy of the original state
which embodies the entanglement between A and B to some extent. This can be
easily understood if the original state is a maximally entangled state (e.g.
a Bell state with $-1$ original conditional entropy and vanishing
post-measurement conditional entropy) \cite{tiaojiu,tiaojiu0}. In addition,
one can find that, if $H(A|B)>0$, the measurement uncertainty is larger than
its disturbance on the state; if $H(A|B)<0$, the measurement uncertainty is
less than its disturbance on the state; and they are equal for $H(A|B)=0$.
In other words, they strongly depend on $H(A|B).$

Now let's turn back to the game with two incompatible observables $\Pi ^{1}$
or $\Pi ^{2}$, the total measurement uncertainties given by $H(\Pi ^{1}|B)+$
$H(\Pi ^{2}|B)$ which, as we know, are bounded by the inequality (3) with X
and Y replaced by $\Pi ^{1}$ and $\Pi ^{2}$ and $c=\max_{k,l}\left\vert
\left\langle \pi _{k}^{1}|\pi _{l}^{2}\right\rangle \right\vert ^{2}$. This
is the M-M relation. Here we would like to find the bound on the measurement
uncertainty of one observable and the disturbance induced by the other
observable, i.e., the M-D relation; and the bound on the disturbances
induced by two incompatible observables, i.e., the D-D relation. Based on
our theorem.1 and Eq. (3), one can easily find the following therorem.

\textit{Theorem}.2. For the projective measurements of the two observables $%
\Pi ^{1}$ and $\Pi ^{2}$ on quantum state $\rho _{AB}$, in the presence of
the quantum memory, the M-D relation is given by
\begin{equation}
H\left( \rho _{AB}\left\Vert \rho _{\Pi ^{1}B}\right. \right) +H\left( \Pi
^{2}|B\right) \geqslant -\log c,  \label{th11}
\end{equation}%
or
\begin{equation}
H\left( \rho _{AB}\left\Vert \rho _{\Pi ^{2}B}\right. \right) +H\left( \Pi
^{1}|B\right) \geqslant -\log c,  \label{th12}
\end{equation}%
and the D-D relation is given by
\begin{equation}
H\left( \rho _{AB}\left\Vert \rho _{\Pi ^{1}B}\right. \right) +H\left( \rho
_{AB}\left\Vert \rho _{\Pi ^{2}B}\right. \right) \geqslant -\log c-H(A|B),
\label{th13}
\end{equation}%
with $c=\max_{k,l}\left\vert \left\langle \pi _{k}^{1}|\pi
_{l}^{2}\right\rangle \right\vert ^{2}.$

\textit{Proof}. Omitted.$\hfill \blacksquare $

From the above theorem, the relationship among the M-M, M-D and D-D
relations is illustrated in Fig. 1 (a). It shows that the three relations
can be converted into each other by considering the contribution of $H(A|B)$%
. But we should note that they reveal different physics.

To demonstrate the relationship among these three relations, we take the
Werner state as the measured state to be an example. The Werner state is
given by \cite{wernerstate}
\begin{equation}
\rho _{AB}=\eta \left\vert \psi ^{\dag }\right\rangle \left\langle \psi
^{\dag }\right\vert +\frac{1-\eta }{4}\mathbb{I},
\end{equation}%
with $\left\vert \psi ^{\dag }\right\rangle =\frac{1}{\sqrt{2}}(\left\vert
00\right\rangle +\left\vert 11\right\rangle )$ the maximally entangled state
and the purity denoted by $\eta $, $0\leqslant \eta \leqslant 1.$ The two
incompatible observables $\Pi ^{1}$ and $\Pi ^{2}$ are performed on
subsystem A. The eigenvectors of $\Pi ^{1}$ is given by
\begin{equation}
\Pi ^{1}:\left\{ \left( \cos \frac{\theta }{2},-e^{i\phi }\sin \frac{\theta
}{2}\right) ,\left( e^{-i\phi }\sin \frac{\theta }{2},\cos \frac{\theta }{2}%
\right) \right\} .  \label{ceX}
\end{equation}%
with the azimuthal angle $0\leqslant \phi \leqslant 2\pi $ and the polar
angle $0\leqslant \theta \leqslant \pi $. Similarly, the other projective
measurement related to ${\Pi ^{2}}$ is defined by
\begin{equation}
\Pi ^{2}:\left\{ \left( \frac{1}{2},\frac{\sqrt{3}}{2}\right) ,\left( \frac{%
\sqrt{3}}{2},-\frac{1}{2}\right) \right\} .  \label{ceY}
\end{equation}%
A straightforward computation gives the quantum conditional entropy of the
initial state $\rho _{AB}$
\begin{equation}
H(A|B)=-\frac{3(1-\eta )}{4}\log \frac{(1-\eta )}{4}-\frac{(1+3\eta )}{4}%
\log \frac{(1+3\eta )}{4}-1.  \label{19}
\end{equation}%
So the total measurement uncertainty (of $\Pi ^{1}$ and $\Pi ^{2}$) reads
\begin{equation}
H(\Pi ^{1}|B)+H(\Pi ^{2}|B)=2-\sum_{m=\pm 1}(1+m\eta )\log (1+m\eta ),
\end{equation}%
the one uncertainty plus the other disturbance is
\begin{eqnarray}
&&H\left( \rho _{AB}\left\Vert \rho _{\Pi ^{1}B}\right. \right) +H(\Pi
^{2}|B)  \notag \\
&=&1+\frac{1+3\eta }{4}\log (1+3\eta )-\frac{1-\eta }{4}\log (1-\eta )
\notag \\
&-&(1+\eta )\log (1+\eta )=H\left( \rho _{AB}\left\Vert \rho _{\Pi
^{2}B}\right. \right) +H(\Pi ^{1}|B),
\end{eqnarray}%
and the total disturbance is%
\begin{eqnarray}
&&H\left( \rho _{AB}\left\Vert \rho _{\Pi ^{1}B}\right. \right) +H\left(
\rho _{AB}\left\Vert \rho _{\Pi ^{2}B}\right. \right)  \notag \\
&=&\frac{1+3\eta }{2}\log (1+3\eta )+\frac{1-\eta }{2}\log (1-\eta )-(1+\eta
)\log (1+\eta ).  \notag \\
\end{eqnarray}%
\begin{figure}[tbp]
\centering
\subfigure[]{\includegraphics[width=1.68in]{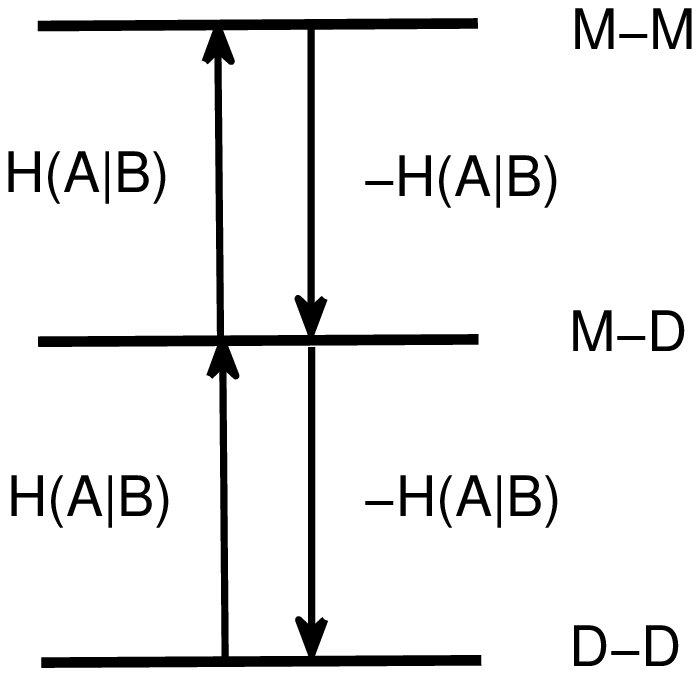}\label{a}} %
\subfigure[]{\includegraphics[width=1.68in]{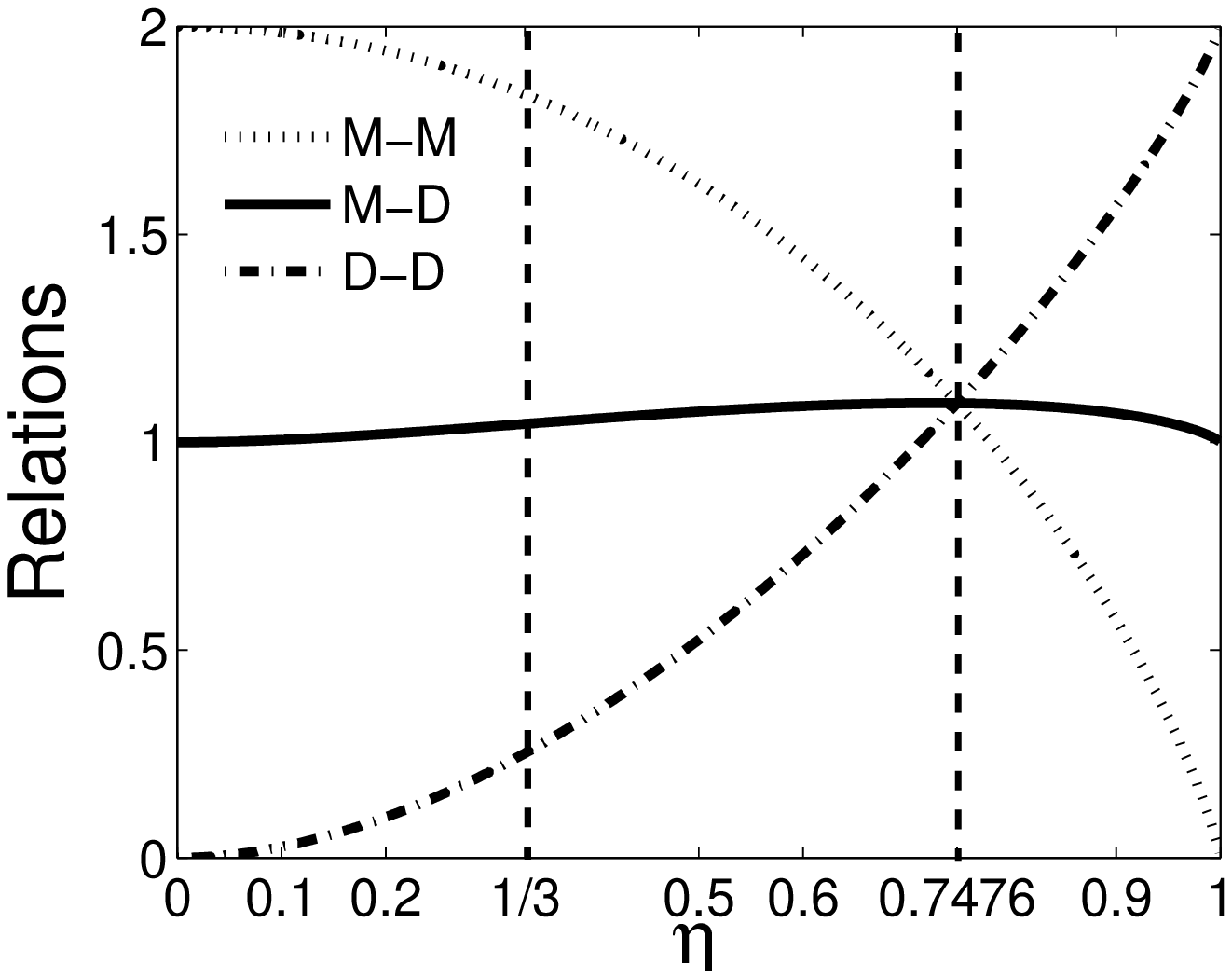}\label{b}}
\caption{(a) The relationship among the M-M, M-D, D-D in the presence of
quantum memory. (b) M-M, M-D, D-D  vs. $\eta$. For the projective measurements the azimuthal angle $\protect\phi =0$
and the polar angle $\protect\theta =\protect\pi /2$, that is, $\Pi
^{1}:\left\{ \left( \frac{1}{\protect\sqrt{2}},-\frac{1}{\protect\sqrt{2}}%
\right) ,\left( \frac{1}{\protect\sqrt{2}},\frac{1}{\protect\sqrt{2}}\right)
\right\} $. The dot line stands for the M-M, the line stands for the M-D and
the dash-dot line stands for the D-D.}
\end{figure}
In Fig.1 (b), we plot the results of Eqs. (13-15). It presents that $H(\Pi
^{1}|B)+H(\Pi ^{2}|B)\geq H\left( \rho _{AB}\left\Vert \rho _{\Pi
^{1}B}\right. \right) +H(\Pi ^{2}|B)\geq H\left( \rho _{AB}\left\Vert \rho
_{\Pi ^{1}B}\right. \right) +H\left( \rho _{AB}\left\Vert \rho _{\Pi
^{2}B}\right. \right) $ below $\eta \approx 0.7476$; they are equal at the
critical point $\eta \approx 0.7476;$ and the relations will be reverse if $%
\eta $ is beyond this critical point.

\section{Uncertainty relations in the absence of quantum memory}

When there is not any quantum memory, that means that the subsystems A and B
have nothing with each other. So we can set $\tilde{\rho}_{AB}=\rho \otimes
\rho _{B}$. Substitute $\tilde{\rho}_{AB}$ into Eq. (5), one will
immediately obtain that
\begin{equation}
H(\rho ||\rho _{\Pi ^{j}})=H(\rho _{\Pi ^{j}})-H(\rho ).  \label{mei}
\end{equation}%
Eq. (\ref{mei}) shows that the measurement uncertainty includes two parts.
One is the disturbance induced by the measurement, the other is the
uncertainty of the measured state itself. From a different angle, one can
also find that the disturbance of a measurement is just the entropy
increment of the post- and pre- measurement states. Since $H(\rho )\geq 0$
for any $\rho $, an important conclusion is that the disturbance by a
measurement is never less than its measurement uncertainty. They are equal
if and only if $\rho $ is a pure state.

If we substitute Eq. (\ref{mei}) into Eqs. (\ref{th11}), (\ref{th12}) and (%
\ref{th13}), one can find that the M-M relation, M-D relation and D-D
relation in the absence of quantum memory as follows.

\textit{Corollary}.1. For the projective measurements of the two observables
$\Pi ^{1}$ and $\Pi ^{2}$ on quantum state $\rho $ in the absence of quantum
memory, the M-M relation is given by [10]%
\begin{equation}
H(\rho _{\Pi ^{1}})+H(\rho _{\Pi ^{2}})\geqslant -2\log c+H(\rho ),
\end{equation}%
the M-D relation reads
\begin{equation}
H(\rho ||\rho _{\Pi ^{1}})+H(\rho _{\Pi ^{2}})\geqslant -\log c,
\end{equation}%
or
\begin{equation}
H(\rho ||\rho _{\Pi ^{2}})+H(\rho _{\Pi ^{1}})\geqslant -\log c,
\end{equation}%
and the D-D relation is
\begin{equation}
H(\rho ||\rho _{\Pi ^{1}})+H(\rho ||\rho _{\Pi ^{2}})\geqslant -\log
c-H(\rho ),
\end{equation}%
where $c=\max_{k,l}\left\vert \left\langle \pi _{k}^{1}|\pi
_{l}^{2}\right\rangle \right\vert ^{2}.$

\textit{Proof}. Omitted.$\hfill \blacksquare $

Considering Eq. (\ref{mei}), one can also find that the three relations given
in Corollary 1 can be converted into each other. Especially, if we consider
Eq. (\ref{mei}) for both observables, one will trivially obtain
\begin{eqnarray}
&&H(\rho _{\Pi ^{1}})+H(\rho _{\Pi ^{2}})  \notag \\
&=&H(\rho ||\rho _{\Pi ^{1}})+H(\rho _{\Pi ^{2}})+H(\rho )  \notag \\
&=&H(\rho ||\rho _{\Pi ^{1}})+H(\rho ||\rho _{\Pi ^{2}})+2H(\rho ).
\label{no2}
\end{eqnarray}%
This also states that the total disturbances are never less than the
measurement uncertainties. From the viewpoint of Heisenberg uncertainty
principle of view, the incompatible observables cannot be simultaneously
accurately determined since the measurements always lead to the disturbance.

\begin{figure}[tbp]
\includegraphics[width=0.8\columnwidth]{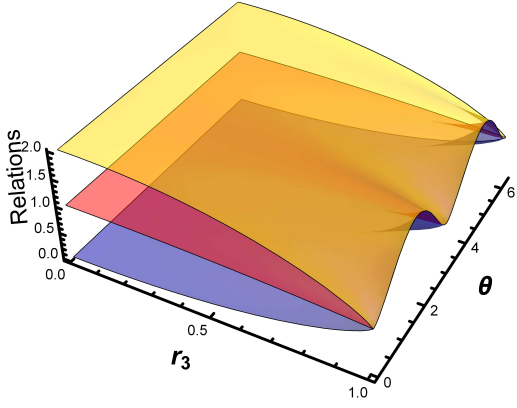}
\caption{(color online) The M-M, the M-D the D-D for the two 2-dimensional projective
measurements vs. $r_{3}$ and $\protect\theta $ in presence of quantum memory. The yellow curved surface stands for the M-M, the red
curved surface stands for the M-D and the blue curved surface stands for the
D-D.}
\end{figure}
As an example, we compare the total uncertainties, the one uncertainty plus
the other disturbance and the total disturbances in the absence of quantum
memory. First, we choose the arbitrary single qubit $\rho =\frac{1}{2}(%
\mathbb{I}+\overrightarrow{r}\cdot \overrightarrow{\sigma })$ with the
standard Pauli operators $\overrightarrow{\sigma }=\{\sigma _{x},\sigma
_{y},\sigma _{z}\}$ and $\sum_{i}r_{i}^{2}=1$ as the measured state. The
incompatible observables $X$ and $Y$ can be taken the same as Eqs. (\ref{ceX}%
,\ref{ceY}). Through a simple calculation, we can obtain the following
expressions:
\begin{eqnarray}
&&H(\rho _{X})+H(\rho _{Y})  \notag \\
&=&\sum_{m=\pm 1}-\frac{1+mr_{3}\cos \theta }{2}\log \frac{1+mr_{3}\cos
\theta }{2}  \notag \\
&&-\frac{2+mr_{3}}{4}\log \frac{2+mr_{3}}{4},
\end{eqnarray}
\begin{eqnarray}
&&H(\rho ||\rho _{X})+H(\rho _{Y})  \notag \\
&=&\sum_{m=\pm 1}\frac{1+mr_{3}}{2}\log \frac{1+mr_{3}}{2}-\frac{2+mr_{3}}{4}%
\log \frac{2+mr_{3}}{4}  \notag \\
&&-\frac{1+mr_{3}\cos \theta }{2}\log \frac{1+mr_{3}\cos \theta }{2},
\end{eqnarray}%
and
\begin{eqnarray}
&&H(\rho ||\rho _{X})+H(\rho ||\rho _{Y})  \notag \\
&=&\sum_{m=\pm 1}(1+mr_{3})\log \frac{1+mr_{3}}{2}-\frac{2+mr_{3}}{4}\log
\frac{2+mr_{3}}{4}  \notag \\
&&-\frac{1+mr_{3}\cos \theta }{2}\log \frac{1+mr_{3}\cos \theta }{2}.
\end{eqnarray}

In Fig.2, we plot the results of Eqs. (22-24). It shows three layers which
stand for our mentioned relations. The upper curved surface (yellow) stands
for the M-M whilst the lower one (blue) is the D-D. In between, the curved surface is M-D. If $r_{3}=1$, the initial quantum state
reduce to a pure state $\rho=|0\rangle\langle0|$, i.e., the von Neumman
entropy is zero, $H(\rho)=0$, they are well consistent.

\section{Uncertainty relations for the multiple measurements in the presence
(and absence) of the quantum memory}

In this section, we extend the M-D relation and the D-D relation to the
multiple measurements in the presence (and absence) of quantum memory. In order to give
a clear background, we would like to describe the game at first. Suppose that Alice performs a group of measurements $\{\Pi
_{i},i=1,2,...,N\}$ on a state $\rho_{AB}$. Alice chooses one measurement $\Pi _{i}$ and announces her choice to
Bob. Bob tries to minimize his uncertainty about Alice's measurement
outcomes. During this game, for the multiple measurements acted on the
initial quantum state, $\{\Pi ^{i},i=1,2,\cdots ,N\}$ can be rearranged in different
orders with $\varepsilon $ labelling the different orders. Thus, $\Pi ^{\varepsilon
_{i}}$ can be understood as \textit{i}th measurement in the $\varepsilon $
order. Similarly, the $\alpha $th eigenvector of $\Pi ^{\varepsilon _{i}}$
can be written as $\left\vert \varepsilon _{i}^{\alpha }\right\rangle $.
With all the above knowledge, the state-dependent entropic uncertainty
relation will be given by \cite{wo}.
\begin{equation}
\sum_{i=1}^{N}H\left( \Pi _{i}|B\right) \geq \max \left\{ \mathcal{L}_{1},%
\mathcal{L}_{opt},0\right\} .  \label{Total}
\end{equation}%
with
\begin{eqnarray}
\mathcal{L}_{1} &=&(N-1)H(A|B)+\max_{\varepsilon }\left\{ \ell _{\varepsilon
}^{U}\right\} , \\
\mathcal{L}_{opt} &=&\frac{N}{2}H\left( A|B\right) +\max_{all\text{ }ways}%
\mathcal{B}_{ways}^{\prime },
\end{eqnarray}%
where
\begin{equation}
\ell _{\varepsilon }^{U}=-\sum\limits_{\alpha _{N}}p_{\varepsilon
_{N}^{\alpha _{N}}}\log \sum\limits_{\alpha _{k},N\geqslant k>1}\max_{\alpha
_{1}}\prod\limits_{n=1}^{N-1}\left\vert \left\langle \varepsilon
_{n}^{\alpha _{n}}|\varepsilon _{n+1}^{\alpha _{n+1}}\right\rangle
\right\vert ^{2},  \label{27}
\end{equation}%
with $p_{\varepsilon _{N}^{\alpha }}=Tr\left( \left\vert \varepsilon
_{N}^{\alpha }\right\rangle \left\langle \varepsilon _{N}^{\alpha
}\right\vert \otimes \mathbf{I}\right) \rho _{AB}$ and $\mathcal{B}%
_{ways}^{\prime }$ is average value of $-\sum\limits_{\alpha
_{2}}p_{\varepsilon _{2}^{\alpha _{2}}}\log \max_{\alpha _{1}}\left\vert
\left\langle \varepsilon _{1}^{\alpha _{1}}|\varepsilon _{2}^{\alpha
_{2}}\right\rangle \right\vert ^{2}$ for all potential two-measurement
combinations, that is $\ell _{\varepsilon }^{U}$ is constrained for only two
measurements.

Suppose that the measurement uncertainties come from $\gamma$ measurements
while the disturbances come from $\beta$ measurements with $\gamma +\beta =N$. Then, the M-D relation and the D-D
relation for the multiple measurements in the presence (and absence) of
quantum memory will be given as follows.

\textit{Corollary.}2. In the presence of quantum
memory, let the initial quantum state $\rho _{AB}$ be measured by the
set of observables $\{\Pi ^{i},i=1,2,\cdots ,N\}$. The state-dependent trade-off relation between
the $\gamma $ measurements uncertainties and the $\beta $ disturbances is given by
\begin{equation}
\gamma \sum\limits_{i=1}^{\gamma }H(\Pi ^{i}|B)+\beta
\sum\limits_{j=1}^{\beta }H(\rho _{AB}||\rho _{\Pi ^{j}B})\geqslant \max
\left\{ \mathcal{L^{\prime }}_{1},\mathcal{L^{\prime }}_{opt},0\right\} ,
\label{35}
\end{equation}%
with
\begin{eqnarray}
\mathcal{L^{\prime }}_{1} &=&(N-1-\beta )H(A|B)+\max_{\varepsilon }\left\{
\ell _{\varepsilon }^{U}\right\} , \\
\mathcal{L^{\prime }}_{opt} &=&(\frac{N}{2}H-\beta )\left( A|B\right)
+\max_{all\text{ }ways}\mathcal{B}_{ways}^{\prime }.
\end{eqnarray}%
The state-dependent trade-off relation in the absence of quantum memory is
given by
\begin{equation}
\gamma \sum\limits_{i=1}^{\gamma }H(\rho _{\Pi ^{i}})+\beta
\sum\limits_{j=1}^{\beta }H(\rho ||\rho _{\Pi ^{j}})\geqslant \max \left\{
\mathcal{L^{\prime \prime }}_{1},\mathcal{L^{\prime \prime }}%
_{opt},0\right\} ,  \label{36}
\end{equation}%
with
\begin{eqnarray}
\mathcal{L^{\prime \prime }}_{1} &=&(N-1-\beta )H(\rho )+\max_{\varepsilon
}\left\{ \ell _{\varepsilon }^{U}\right\} , \\
\mathcal{L^{\prime \prime }}_{opt} &=&(\frac{N}{2}H-\beta )H(\rho )+\max_{all%
\text{ }ways}\mathcal{B}_{ways}^{\prime },
\end{eqnarray}%
where $\ell _{\varepsilon }^{U}$ is given by Eq. (\ref{27}).

\textit{Proof}. Omitted$\hfill \blacksquare $

It is obvious that the term $\ell _{\varepsilon }^{U}$ which describes the
complementarity of the non-degenerate observables depends on the sequence of
observables and the initial quantum state. In order to eliminate the
state dependency, we take maximum over $\alpha _{N}$ of $\Pi ^{\varepsilon
_{N}}$, so $\ell _{\varepsilon }^{U}$ in the second term becomes
\begin{eqnarray}
\ell _{\varepsilon }^{U} &=&-\sum\limits_{\alpha _{N}}p_{\varepsilon
_{N}^{\alpha _{N}}}\log \sum\limits_{\alpha _{k},N\geqslant k>1}\max_{\alpha
_{1}}\prod\limits_{n=1}^{N-1}\left\vert \left\langle \varepsilon
_{n}^{\alpha _{n}}|\varepsilon _{n+1}^{\alpha _{n+1}}\right\rangle
\right\vert ^{2}  \notag \\
&\geqslant &-\max_{\alpha _{N}}\log \sum\limits_{\alpha
_{k},N>k>1}\max_{\alpha _{1}}\prod\limits_{n=1}^{N-1}\left\vert \left\langle
\varepsilon _{n}^{\alpha _{n}}|\varepsilon _{n+1}^{\alpha
_{n+1}}\right\rangle \right\vert ^{2}=\ell _{\varepsilon }^{\tilde{U}}.
\label{wu1}
\end{eqnarray}%
Replacing $\ell _{\varepsilon }^{U}$ with $\ell _{\varepsilon }^{\tilde{U}}$ 
in Eqs. (\ref{35},\ref{36}), we can obtain the corresponding state-independent
trade-off relation between the $\gamma $ measurements uncertainties and the $\beta
$ disturbances in the
presence (and absence) of quantum memory. From the Corollary.2, we can
choose $\gamma =N,\beta =0$ measurements to establish the M-M relation
whilst choose $\gamma =0,\beta =N$ measurements to establish the D-D
relation. At the same time the relationship among the M-M, the M-D and the
D-D for the multiple measurements is similar to the relationship for two incompatible
measurements.

\section{Conclusion and Discussion}

We present the relation between the measurement uncertainty and its disturbance in terms of quantum relative entropy.
 We find that in the presence of quantum memory, the measurement uncertainty is closely related to its disturbance on the measured state
 by the conditional entropy of the original state, and in the absence of quantum memory, the disturbance of a measurement can be described by its measurement uncertainty plus the uncertainty of the measured state. In other words, the disturbance can be understood by the entropic increment of the post- and pre- measurement state.
 Based on such relations, we study the measurement-measurement uncertainty relation, the
measurement-disturbance relation and the disturbance tradeoff relation in
the presence (and absence) of quantum memory.
At the same time, we also establish the relationship among the M-M, D-D and M-D in the presence (absence) of quantum memory. Our results
have also been extended to the case with multiple measurements.
Finally, we also demonstrate these relations by concrete example.

\acknowledgements{This work was supported by the National Natural Science Foundation of China, under Grant
No.11375036 and 11175033, the Xinghai Scholar Cultivation Plan and the Fundamental
Research Funds for the Central Universities under Grant No. DUT15LK35.}

\end{document}